\documentclass[authoryear,review,11pt,a4paper]{elsarticle}

\usepackage{setspace}
\usepackage{fullpage}

\usepackage{color}

\usepackage{amsmath}

\usepackage{amsfonts}
\usepackage{amssymb}
\usepackage{amsthm}

\usepackage{amscd}

\usepackage{mathtools}

\usepackage{bbm}

\usepackage{tikz}
\usepackage{tikz-cd}

\usepackage[ruled]{algorithm2e}

\usepackage{textgreek}

\usepackage{graphicx}

\usepackage{lscape}
\usepackage{afterpage}

\usepackage{booktabs}
\usepackage{multirow}
\usepackage{float}
\usepackage{caption}
\usepackage{subcaption}
\usepackage{longtable}

\usepackage{lineno}

\usepackage{hyperref}

\biboptions{authoryear}

\newdefinition{rmk}{Remark}
\newproof{pf}{Proof}

\journal{Statistics in Medicine}

\begin{document}

\begin{frontmatter}



\title{A Bayesian model-free dose-finding approach in Phase II clinical trials with the flexibility of historical borrowing}

\author[BrisMath]{Linxi Han\fnref{fn3}}

\author[Moderna]{Qiqi Deng}


\author[BrisMath]{Feng Yu\corref{cor1}}
\ead{feng.yu@bristol.ac.uk}

\author[TGIUK,ICTU]{Zhangyi He\corref{cor1}\fnref{fn3}}
\ead{zhe1@georgeinstitute.org.uk}

\cortext[cor1]{Corresponding author.}
\fntext[fn3]{Affiliation at time of study}

\address[BrisMath]{School of Mathematics, University of Bristol, Bristol, UK}
\address[Moderna]{Biostatistics, Moderna Inc, Cambridge, Connecticut, USA}
\address[TGIUK]{The George Institute for Global Health, Imperial College London, London, UK}
\address[ICTU]{Imperial Clinical Trials Unit, Imperial College London, London, UK}

\begin{abstract}
Accurate dose selection in Phase II trials is critical to the success of subsequent Phase III trials, but suboptimal choices remain a leading cause of trial failure and regulatory rejection. Although MCP-Mod is widely adopted and endorsed by regulatory agencies, it requires prespecification of candidate models and is highly sensitive to model misspecification. To address these challenges, we introduce MAP-curvature, a general model-free framework for dose-response modelling that penalises the total curvature of the dose-response curve through a prior. Within this framework, LiMAP-curvature arises as the linear special case, whereas SEMAP-curvature, the focus of this work, employs the sigmoid Emax model, providing greater flexibility to capture nonlinear pharmacological patterns. Through extensive simulations, we show that SEMAP-curvature generally outperforms LiMAP-curvature and MCP-Mod in detecting the dose-response signal, estimating the dose-response curve and identifying the minimum effective dose, with particularly significant improvements under concave downward shapes resembling the sigmoid Emax model. Although SEMAP-curvature exhibits slightly greater variability, it remains robust in accuracy and reliability. We further extend MAP-curvature by integrating it with the Bayesian hierarchical model to enable flexible borrowing of historical data, which improves power and precision, particularly when dose levels overlap across studies. These results highlight MAP-curvature, and in particular SEMAP-curvature with historical borrowing, as a robust and efficient framework for dose selection in early-phase clinical trials.
\end{abstract}


\end{frontmatter}


\section{Introduction}
\label{sec:1}
The development of a new pharmaceutical product is a lengthy and costly process, commonly taking an average of eight years from clinical trials to market approval \citep{kaitin2011}. A major challenge within this process is dose selection in Phase II trials, which is critical to the success of subsequent Phase III pivotal trials. Suboptimal dose selection has been identified as a primary driver of Phase III trial failures, accounting for nearly half of such cases, and regulatory rejections are frequently driven not by inadequate efficacy or safety, but by insufficient evidence to support dose-response relationships.

A review by \citet{sacks2014} on New Molecular Entities submitted to the FDA's Center for Drug Evaluation and Research found that uncertainty in dose selection was the main reason for failure in initial drug applications. This issue, defined by the inability to determine the optimal dose that maximises therapeutic benefits while minimising safety risks, underscores the critical importance of accurately understanding dose-response relationships, which are fundamental to the success of new drug applications and represent one of the most complicated and challenging aspects of the clinical development \citep{chen2020}. As noted by \cite{bretz2005}, selecting a dose that is too high may result in severe safety concerns and unacceptable adverse events in later phases, while selecting a dose that is too low may fail to deliver adequate clinical benefits. Thus, efficient and reliable methods for dose-response evaluation are essential, particularly those that maximise information from limited trial data.

Traditionally, dose-finding trials have been analysed through two distinct approaches: Multiple Comparison Procedures (MCP) and Modelling (Mod) techniques. \citet{bretz2005} combined these two methods into a unified framework known as Multiple Comparison Procedures-Modelling (MCP-Mod), which was later generalised and improved by \citet{pinheiro2014}. This procedure has been approved by regulatory authorities \citep{EMA2014,FDA2016} and has gained widespread popularity in Phase II trials over the past decade \citep[\textit{e.g.},][]{shah2019,yazici2021,bowman2022}, valued for its ability to combine the strengths of MCP and Mod to address uncertainties in dose-response relationships. While MCP-Mod provides an efficient and robust statistical framework for dose-finding studies, its implementation can be complex. MCP-Mod requires the pre-specification of candidate models with their parameters, which can be challenging as prior knowledge about the agent is usually limited \citep{chen2020}. Model mis-specification can significantly reduce the statistical power of the analysis \citep{saha2019}. Additionally, MCP-Mod encounters convergence issues, particularly when it is used in combination with other complicated models \citep{Zhang2024}. Alternative approaches for dose-finding trials are reviewed by \citet{ting2006}.

To overcome these limitations, \cite{han2022} introduced LiMAP-curvature, a model-free Bayesian approach that detects dose-response signals without relying on pre-specified candidate models, and demonstrated its superiority over MCP-Mod, especially in scenarios where the true model greatly deviates from the candidate model set specified in MCP-Mod. LiMAP-curvature is based on an $L^2$-total curvature, which measures how far the true dose-response curve deviates from linearity. This curvature term imposes a certain degree of smoothness on the dose-response relationship. Although linear dose-response relationships may arise in simple mechanisms or at low dose levels, they are generally uncommon in pharmacology \citep{tsatsakis2018}.

Building on the LiMAP-curvature, we propose a MAP-curvature framework, a general dose-finding approach that enables flexible modelling of dose-response relationships by incorporating the total (in the $L^2$ sense) curvature of the dose-response curve as a prior parameter. Responses at the prespecified dose levels are estimated through maximum \textit{a posteriori} (MAP), from which we construct a test statistic to establish proof of concept (PoC) by simulation. LiMAP-curvature arises as a special case of this framework, corresponding to a linear default dose-response function. While effective under approximately linear relationships, LiMAP-curvature loses efficiency under more complex nonlinear patterns, where a straight line may fail to capture the underlying trend. To address this limitation, we introduce SEMAP-curvature, a variant of MAP-curvature that adopts the sigmoid Emax model as the default dose-response function. The sigmoid Emax model is widely recognised for its ability to approximate the monotonic nonlinear dose-response curves commonly observed in pharmacological studies \citep{ting2006}. By leveraging this model, SEMAP-curvature increases the accuracy of dose-response estimation and offers a more robust framework for dose selection in early-phase trials, especially when the true relationship deviates from linearity.

A persistent challenge in dose-finding studies is the limited data available for accurate dose estimation. \citet{neuenschwander2010} highlighted that historical trial data can substantially inform the design and analysis of new trials. Incorporating such information may shorten trial duration or reduce the sample size needed to achieve adequate power, therefore reducing overall development cost \citep{schmidli2014}. As a result, there has been growing interest in methods that formally borrow strength from historical data in the design and analysis of dose-finding trials \citep[\textit{e.g.},][]{fleischer2022,han2024}. Here we further extend the MAP curvature framework by integrating it with the Bayesian hierarchical model proposed by \citet{han2024} to allow flexible historical borrowing in dose-finding trials. This approach combines the strengths of MAP-curvature for modelling complex dose-response relationships with the advantages of the Bayesian hierarchical model, which allows historical information to be incorporated across an arbitrary number of dose groups while accounting for prognostic and predictive effects between trials. By dynamically adjusting the degree of borrowing based on the homogeneity in treatment effects across studies, our approach improves dose estimation in current trials while addressing between-trial heterogeneity. This combined strategy circumvents key challenges in dose-finding studies, such as small sample sizes, limited prior knowledge and complex dose-response patterns, thereby supporting more robust and efficient dose selection.

The remainder of this work is structured as follows: Section~\ref{sec:2} introduces the MAP-curvature framework, along with its extension to incorporate historical borrowing and a detailed discussion of SEMAP-curvature. Section~\ref{sec:3} investigates the operating characteristics of SEMAP-curvature across various scenarios, both with and without borrowing historical data. Section~\ref{sec:4} concludes with a summary of the findings and explores potential directions for future research.

\section{Materials and Methods}
\label{sec:2}
In this section, we first provide a theoretical introduction of the MAP-curvature framework, followed by a brief review of LiMAP-curvature and a detailed description of SEMAP-curvature. 
Moreover, we demonstrate how to extend the MAP-curvature framework to integrate historical data when available.

\subsection{MAP-curvature}
\label{sec:21}
We consider a trial with a total of $M+1$ parallel groups of patients corresponding to doses $x_{1},x_{2},\ldots,x_{M}$ plus placebo $x_{0}$. Similar to \citet{bretz2005}, we let $Y_{ij} \mid \mu_{i} \sim N(\mu_{i},\sigma^{2})$ be the response of patient $j=1,2,\ldots,N_{i}$ at dose $x_{i}$, where $\mu_{i}=f(x_{i})$ represents the mean response at dose $x_{i}$ for some dose-response model $f(x)$ defined over $[0,1]$ and $N(\mu,\sigma^{2})$ denotes the normal distribution with mean $\mu$ and variance $\sigma^{2}$. We assume that $\sigma^{2}$ is known and constant across all dose groups. 

In LiMAP-curvature, \citet{han2022} introduced an $L^2$-total curvature, defined by
\begin{linenomath}
	\begin{equation}
		\label{eqn:2102}
		Sf
		=
		\left(\int_{x_{0}}^{x_{M}} \left(\frac{df}{dx^{2}}(x)\right)^{2} \, dx \right)^{1/2},
	\end{equation}
\end{linenomath}
to measure the deviation of the true dose-response curve from a straight line, and this curvature term guarantees that the model enforces a degree of smoothness on the true dose-response curve. However, a linear dose-response curve may not serve as an appropriate default model since such relationships are rarely observed in practice \citep{tsatsakis2018}. To address this limitation, we generalise the LiMAP-curvature framework to accommodate other, more commonly observed dose-response curves as default models. 

We let $\phi(x;\boldsymbol{\theta})$ be the default dose-response model with its corresponding parameters $\boldsymbol{\theta}$, and then we have $(\phi^{-1} \circ \phi)(x;\boldsymbol{\theta})=x$, where $\phi^{-1}$ denotes the inverse of $\phi$. With Eq.~(\ref{eqn:2102}), we re-define the $L^2$-total curvature as
\begin{linenomath}
	\begin{equation}
		\label{eqn:2103}
		S\phi^{-1}(f)
		=
		\left(\int_{x_{0}}^{x_{M}} \left(\frac{d\phi^{-1}}{dx^{2}}(f(x);\boldsymbol{\theta})\right)^{2} \, dx \right)^{1/2},
	\end{equation}
\end{linenomath}
quantifying how far the true dose-response curve $f(x)$ deviates from the default dose-response model $\phi(x;\boldsymbol{\theta})$. Higher values of $S\phi^{-1}(f)$ indicate greater irregularity in the dose-response curve $f(x)$, reflecting deviations from the default model $\phi(x;\boldsymbol{\theta})$. Conversely, lower values of $S\phi^{-1}(f)$ suggest a smoother dose-response curve that closely aligns with the default model.

Following \citet{han2022}, we impose a half-normal prior distribution, $HN(\gamma^2)$, on $S\phi^{-1}(f)$ to balance the trade-off between the smoothness of the dose-response curve and its fidelity to the observed data. Such a prior favours smoother dose-response curves, with $\gamma$ serving as a tuning parameter to adjust the balance between encouraging smoothness and allowing deviations from the default model. We place a half-normal hyperprior distribution, $HN(\tau^2)$, on $\gamma$, incorporating additional uncertainty into the prior of $S\phi^{-1}(f)$, to further improve its flexibility.

Through the second-order central difference scheme \citep{burden2015}, we have
\begin{linenomath}
	\begin{equation}
	\label{eqn:2104}
		\frac{d\phi^{-1}}{dx^{2}}(f(x_{i});\boldsymbol{\theta})
		\approx
		2\left(\frac{\phi^{-1}(\mu_{i+1};\boldsymbol{\theta})-\phi^{-1}(\mu_{i};\boldsymbol{\theta})}{(x_{i+1}-x_{i})(x_{i+1}-x_{i-1})}-\frac{\phi^{-1}(\mu_{i};\boldsymbol{\theta})-\phi^{-1}(\mu_{i-1};\boldsymbol{\theta})}{(x_{i}-x_{i-1})(x_{i+1}-x_{i-1})}\right)
	\end{equation}
\end{linenomath}
for $i=1,2,\ldots,M-1$. With Eq.~(\ref{eqn:2104}), we can then approximate the $L^2$-total curvature $S\phi^{-1}(f)$ in Eq.~(\ref{eqn:2103}) through numerical integration as
\begin{linenomath}
	\begin{equation}
		\label{eqn:2105}
		S_{\boldsymbol\mu}
		=
		2 \left(\sum_{i=1}^{M-1} \left(\frac{\phi^{-1}(\mu_{i+1};\boldsymbol{\theta})-\phi^{-1}(\mu_{i};\boldsymbol{\theta})}{(x_{i+1}-x_{i})(x_{i+1}-x_{i-1})}-\frac{\phi^{-1}(\mu_{i};\boldsymbol{\theta})-\phi^{-1}(\mu_{i-1};\boldsymbol{\theta})}{(x_{i}-x_{i-1})(x_{i+1}-x_{i-1})}\right)^{2} \Delta x_{i}\right)^{1/2},
	\end{equation}
\end{linenomath}
where $\Delta x_{i}=(x_{i+1}-x_{i-1})/2$ for $i=2,3,\ldots,M-2$, along with $\Delta x_{1}=(x_{2}+x_{1})/2-x_{0}$ and $\Delta x_{M-1}=x_{M}-(x_{M-1}+x_{M-2})/2$. 

We let $\boldsymbol{Y}=\{Y_{ij} \mid i=0,1,\ldots,M, j=1,2,\ldots,N_{i}\}$ represent all available patient response data from the trial. Combining with Eq.~(\ref{eqn:2105}), we can formulate the posterior for $\boldsymbol\mu$, $\gamma$ and $\boldsymbol{\theta}$ as
\begin{linenomath}
	\begin{equation}
		\label{eqn:2106}
		p(\boldsymbol\mu,\gamma,\boldsymbol{\theta} \mid \boldsymbol{Y}) 
		\propto
		p(\boldsymbol\mu,\gamma,\boldsymbol{\theta})p(\boldsymbol{Y} \mid \boldsymbol\mu,\gamma,\boldsymbol{\theta})
	\end{equation}
\end{linenomath}
with the prior
\begin{linenomath}
	\begin{equation}
		\label{eqn:2107}
		p(\boldsymbol\mu,\gamma,\boldsymbol{\theta}) 
		=
		p(\gamma) p(\boldsymbol{\theta}) p(S_{\boldsymbol\mu} \mid \gamma,\boldsymbol{\theta}) \prod_{i=0}^{M} p(\mu_{i}) 
	\end{equation}
\end{linenomath}
and the likelihood
\begin{linenomath}
	\begin{equation}
		\label{eqn:2108}
		p(\boldsymbol{Y} \mid \boldsymbol\mu,\gamma,\boldsymbol{\theta})
		=
		\prod_{i=0}^{M}\prod_{j=1}^{N_{i}} p(Y_{ij} \mid \mu_{i}),
	\end{equation}
\end{linenomath}
where we assume $\mu_{i} \sim U(0,1)$ for $i=0,1,\ldots,M$ with $U(a,b)$ denoting the uniform distribution over $[0,1]$. The choice of the prior $p(\boldsymbol{\theta})$ depends on the default dose-response model specification. Using Eqs.~(\ref{eqn:2106})--(\ref{eqn:2108}), we can achieve the MAP estimates
\begin{linenomath}
	\begin{align}
		\label{eqn:2109}
		\hat{\boldsymbol\mu},\hat{\gamma},\hat{\boldsymbol{\theta}}
		&=
		\arg\max_{\boldsymbol\mu,\gamma,\boldsymbol{\theta}} \log p(\boldsymbol\mu,\gamma,\boldsymbol{\theta} \mid \boldsymbol{Y}) \notag \\
		&=
		\arg\max_{\boldsymbol\mu,\gamma,\boldsymbol{\theta}} \left( - \frac{\gamma^{2}}{2\tau^{2}} + \log p(\boldsymbol{\theta}) + \log \gamma - \frac{S_{\boldsymbol{\mu}}^{2}}{2\gamma^{2}} - \sum_{i=0}^{M}\sum_{j=1}^{N_{i}} \frac{(Y_{ij}-\mu_{i})^{2}}{2\sigma^{2}} \right)
	\end{align}
\end{linenomath}
through a numerical optimisation approach like the Broyden-Fletcher-Goldfarb-Shanno (BFGS) algorithm or its variants \citep[see, \textit{e.g.},][for more details]{nocedal1999}. We can estimate the dose-response curve with $\hat{\boldsymbol\mu}$ through linear interpolation (or a more sophisticated interpolation approach), which will also yield an estimate of the target dose of interest.

To establish PoC, we test the null hypothesis $H_{0}:\mu_{1}=\mu_{2}=\ldots=\mu_{M}=\mu_{0}$ against the one-sided alternative hypothesis $H_{1}:\max\{\mu_{1},\mu_{2},\ldots,\mu_{M}\}>{\mu_{0}}$. We introduce a test statistic $T=\max\{\mu_{1},\mu_{2},\ldots,\mu_{M}\}-\mu_{0}$, and given a significance level of $\alpha$, the corresponding critical value $c$ is selected such that $\mathbb{P}(T>c \mid H_{0})<\alpha$. The critical value $c$ is calibrated through Monte Carlo simulation with
\begin{linenomath}
	\begin{equation*}
		\Pr(T>c \mid H_{0})
		\approx
		\frac{1}{R}\sum_{r=1}^{R}\mathbbm{1}_{\{T^{(r)}>c\}}
		<
		\alpha,
	\end{equation*}
\end{linenomath}
where $T^{(r)}$ is the test statistic computed with Eq.~(\ref{eqn:2109}) from the data $\boldsymbol{Y}^{(r)}$ simulated under $H_{0}$ for the $r$-th replicate, $R$ is the total number of replicates, and $\mathbbm{1}_{A}$ is the indicator function equal to $1$ if condition $A$ holds and 0 otherwise.

\subsubsection{LiMAP-curvature}
\label{sec:211}
LiMAP-curvature represents a special case of our MAP-curvature framework, in which the default dose-response model is taken to be $\phi(x;\boldsymbol{\theta})=x$ with its inverse $\phi^{-1}(y;\boldsymbol{\theta})=y$. The MAP estimates in Eq.~(\ref{eqn:2109}) therefore reduce to
\begin{linenomath}
	\begin{equation*}
		\hat{\boldsymbol\mu},\hat{\gamma}
		=
		\arg\max_{\boldsymbol\mu,\gamma} \left( - \frac{\gamma^{2}}{2\tau^{2}} + \log \gamma - \frac{S_{\boldsymbol{\mu}}^{2}}{2\gamma^{2}} - \sum_{i=0}^{M}\sum_{j=1}^{N_{i}} \frac{(Y_{ij}-\mu_{i})^{2}}{2\sigma^{2}} \right)
	\end{equation*}
\end{linenomath}
with
\begin{linenomath}
	\begin{equation*}
		S_{\boldsymbol\mu}
		=
		2 \left(\sum_{i=1}^{M-1} \left(\frac{\mu_{i+1}-\mu_{i}}{(x_{i+1}-x_{i})(x_{i+1}-x_{i-1})}-\frac{\mu_{i}-\mu_{i-1}}{(x_{i}-x_{i-1})(x_{i+1}-x_{i-1})}\right)^{2} \Delta x_{i}\right)^{1/2},
	\end{equation*}
\end{linenomath}
which are consistent with \citet{han2022}.
Full details on the implementation and performance of LiMAP-curvature can be found in \citet{han2022}.

\subsubsection{SEMAP-curvature}
\label{sec:212}
While linear dose-response relationships can occur, non-linear curves, like sigmoidal shapes, are more common in pharmacological and toxicological studies due to factors such as receptor saturation and feedback mechanisms \citep{tsatsakis2018}. We therefore introduce SEMAP-curvature, which defines the default dose-response model in the MAP-curvature framework as a sigmoid Emax model \citep{holford1981,ting2006}:
\begin{linenomath}
	\begin{equation*}
		\phi(x;\boldsymbol{\theta})
		=
		E_{0}+E_{\max} \frac{x^\lambda}{x^\lambda+ED_{50}^\lambda}
	\end{equation*}
\end{linenomath}
with its inverse
\begin{linenomath}
	\begin{equation*}
		\phi^{-1}(y;\boldsymbol{\theta})
		=
		ED_{50} \left(\frac{E_{\max}}{E_{\max}+E_{0}-y}-1\right)^{1/\lambda},
	\end{equation*}
\end{linenomath}
where $\boldsymbol{\theta}=\{E_{0},E_{\max},ED_{50},\lambda\}$ with $E_{0}$ denoting the baseline effect, $E_{\max}$ being the maximum achievable effect, $ED_{50}$ denoting the dose achieving 50\% of $E_{\max}$ and $\lambda$ being the hill coefficient controlling the steepness of the dose-response curve. 
The sigmoid Emax model is usually used in practice due to its ability to approximate a wide range of monotonic dose-response relationships \citep{thomas2006}. 

We specify $p(\boldsymbol{\theta})$ based on historical data, pharmacokinetic information and broader insights from previous studies on dose-response relationships \citep{thomas2006}. More specifically, we pick normal prior distributions for $E_{0}$ and $E_{\max}$, where the means are informed by historical evidence or expert opinion when available, with small variances to reflect high certainty. When such prior knowledge is unavailable, large variances are assigned to accommodate greater uncertainty. As reported by \cite{dutta1996}, estimated values of $\lambda$ near 2 are most commonly observed across a large set of compounds they reviewed, while values above 5 or below 1 are rare. Accordingly, we pick a gamma, beta or log-normal prior distribution for $\lambda$ \citep{thomas2006,wu2018}, which allows flexibility in capturing plausible values while assigning low probabilities to extreme values outside this typical range. For $ED_{50}$, we choose a normal prior distribution truncated to $[0,1]$, consistent with the normalisation of doses between 0 and 1. 

The remainder of SEMAP-curvature adhere to the MAP-curvature framework outlined in Section \ref{sec:21}. 

\subsection{Historical borrowing}
\label{sec:22}
To extend our MAP-curvature framework for historical borrowing, we let $x_{0},x_{1},\ldots,x_{M}$ be all doses (including the placebo $x_{0}$) tested in either the current trial or the historical trial. The sets $\mathcal{I}^{(c)}$ and $\mathcal{I}^{(h)}$ denote the indices of doses used in the current and historical trials, respectively. Incorporating historical data into the current clinical trial necessitates careful adjustments for potential differences between the trials due to prognostic and/or predictive heterogeneity. 

Following \citet{han2024}, we model the response of patient $j=1,2,\ldots,N_{i}^{(c)}$ at dose $x_{i}$ in the current trial as $Y_{i j}^{(c)} \mid \mu_{i},r \sim N(\mu_{i}+r,\sigma^{2})$, where $i \in \mathcal{I}^{(c)}$. Similarly, for the historical trial, the response of patient $j=1,2,\ldots,N_{i}^{(h)}$ at dose $x_{i}$ is modelled as $Y_{i j}^{(h)} \mid \mu_{i},a,r \sim N(a\mu_{i}-r,\sigma^{2})$, where $i \in \mathcal{I}^{(h)}$. Here, $r \sim N(0,\rho^{2})$ quantifies the between-trial heterogeneity in the prognostic effect, while $a \sim N_{[b,1/b]}(1,\eta^{2})$ measures the between-trial heterogeneity in the predictive effect, where $N_{[b, 1/b]}$ represents a normal distribution truncated to $[b, 1/b]$ for some $0 <b<1$, with $b=1/3$ often chosen to ensure robustness and reflect realistic constraints \citet{han2024}. 

To incorporate historical data into MAP-curvature while adjusting for between-trial predictive and prognostic heterogeneity, we can represent the posterior for $\boldsymbol\mu$, $\gamma$, $\boldsymbol{\theta}$, $a$ and $r$ as
\begin{linenomath}
	\begin{equation}
		\label{eqn:2201}
		p(\boldsymbol\mu,\gamma,\boldsymbol{\theta},a,r \mid \boldsymbol{Y}^{(c)},\boldsymbol{Y}^{(h)}) 
		\propto
		p(\boldsymbol\mu,\gamma,\boldsymbol{\theta},a,r)p(\boldsymbol{Y}^{(c)},\boldsymbol{Y}^{(h)} \mid \boldsymbol\mu,\gamma,\boldsymbol{\theta},a,r)
	\end{equation}
\end{linenomath}
with the prior
\begin{linenomath}
	\begin{equation*}
		p(\boldsymbol\mu,\gamma,\boldsymbol{\theta},a,r)
		=
		p(a)p(r)p(\gamma) p(\boldsymbol{\theta}) p(S_{\boldsymbol\mu} \mid \gamma,\boldsymbol{\theta}) \prod_{i=0}^{M} p(\mu_{i}) 
	\end{equation*}
\end{linenomath}
and the likelihood
\begin{linenomath}
	\begin{equation*}
		p(\boldsymbol{Y}^{(c)},\boldsymbol{Y}^{(h)} \mid \boldsymbol\mu,\gamma,\boldsymbol{\theta},a,r)
		=
		\left(\prod_{i \in \mathcal{I}^{(c)}}\prod_{j=1}^{N_{i}^{(c)}} p(Y_{ij}^{(c)} \mid \mu_{i},r)\right) \left(\prod_{i \in \mathcal{I}^{(h)}}\prod_{j=1}^{N_{i}^{(h)}} p(Y_{ij}^{(h)} \mid \mu_{i},a,r)\right), 
	\end{equation*}
\end{linenomath}
where $\boldsymbol{Y}^{(c)}=\{Y_{ij}^{(c)} \mid i \in \mathcal{I}^{(c)}, j=1,2,\ldots,N_{i}^{c}\}$ and $\boldsymbol{Y}^{(h)}=\{Y_{ij}^{(h)} \mid i \in \mathcal{I}^{(h)}, j=1,2,\ldots,N_{i}^{(h)}\}$. The posterior in Eq.~(\ref{eqn:2201}) integrates information from both the current and historical trials while adjusting for their differences in baseline characteristics and treatment effects, thereby ensuring robustness and flexibility in the analysis of dose-finding trials. Using an appropriate numerical optimisation algorithm, we can achieve the MAP estimates
\begin{linenomath}
	\begin{equation*}
		\hat{\boldsymbol\mu},\hat{\gamma},\hat{\boldsymbol{\theta}},\hat{a},\hat{r}
		=
		\arg\max_{\boldsymbol\mu,\gamma,\boldsymbol{\theta},a,r} \log p(\boldsymbol\mu,\gamma,\boldsymbol{\theta},a,r \mid \boldsymbol{Y}^{(c)},\boldsymbol{Y}^{(h)}).
	\end{equation*}
\end{linenomath}
The procedures for dose-response signal detection, dose-response curve estimation and minimum effective dose (MED) determination follow those in the MAP-curvature framework described in Section \ref{sec:21}.

\section{Simulation studies}
\label{sec:3}
This section presents a comprehensive series of simulation studies conducted to evaluate the performance of SEMAP-curvature across a broad range of scenarios. The evaluation focuses on key operating characteristics, including its ability to detect the dose-response signal, estimate the dose-response relationship and identify the MED.

\subsection{Setup of the simulation study}
\label{sec:31}
Our simulation studies are based on a randomised, double-blind, placebo-controlled, parallel-group trial with patients being equally allocated to the placebo arm (dose 0) or one of four active arms (doses 0.15, 0.50, 0.80 and 1). We let $\mathcal{D}_{c}=\{0,0.15,0.5,0.8,1\}$ be the set of doses tested in the current trial. The placebo effect is fixed at 0, with the maximum treatment effect set at 0.5. The total sample size is 200, corresponding to 40 patients per arm. Responses are assumed to be normally distributed with standard deviation $\sigma=1$.

We consider four scenarios from \citet{han2022}: three incorporating historical trials and one without, as summarised in Table~\ref{tab:311}. 
In each scenario, the true dose-response relationship is selected from 12 common shapes, as illustrated in Figure~\ref{fig:311} with their corresponding parameters provided in Supplemental Material, Table~\href{run:./LH2023_Supplemental_Material.pdf}{S1}. When historical trials are available, we vary the levels of between-trial heterogeneity in predictive and prognostic effects, with $a \in \{1,0.8\}$ and $r \in \{0,0.2\}$. This setup yields a total of 156 scenario-heterogeneity combinations, and for each, we generate 10,000 virtual trials.

\begin{table}[h]
	\centering
	\begin{tabular}{cccc}
		\toprule
		Scenario & Source                        & $\mathcal{D}_{h}$      & $|\mathcal{D}_{c} \cap \mathcal{D}_{h}|$ \\ 
		\midrule
		1        & Phase I dose escalation trial & $\{0,0.15,0.5,0.8,1\}$ & 5 \\
		2        & Phase I dose escalation trial & $\{0,0.15,0.2,0.8,1\}$ & 4 \\
		3        & Phase IIa PoC trial           & $\{0,0.8,1\}$          & 3 \\
		4        & Not available                 & $\emptyset$            & 0 \\
		\bottomrule
	\end{tabular}
	\caption{Summary of four simulation scenarios, where $\mathcal{D}_{h}$ represents the set of doses tested in the historical trial, and $|\mathcal{D}_{c} \cap \mathcal{D}_{h}|$ represents the number of overlapping doses between the current trial and the historical trial.}
	\label{tab:311}
\end{table}

\begin{figure}[!ht]
	\centering
	\includegraphics[width=\linewidth]{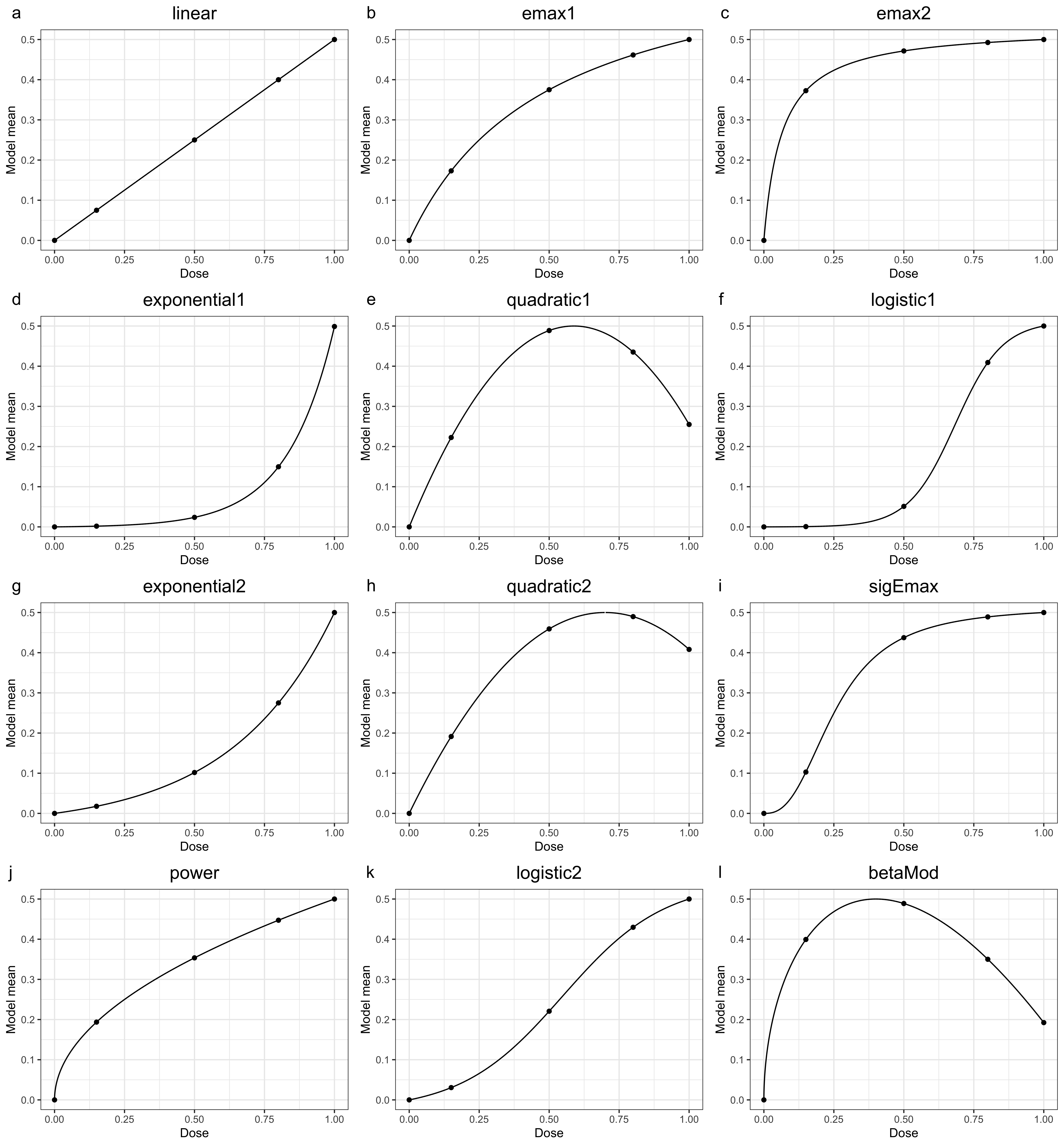}
	\caption{12 common dose-response shapes.} 
	\label{fig:311}
\end{figure}

For each virtual trial, we detect the dose-response signal, estimate the dose-response curve and determine the MED using SEMAP-curvature. In this method, the placebo response is fixed at zero, eliminating the need for the placebo effect $E_{0}$. The prior for the maximum effect $E_{\max}$ is assigned a normal distribution $N(0.5,0.2^2)$, assuming a moderate and well-defined maximum response. The prior for the median effective dose $ED_{50}$ is set to a truncated normal distribution $N_{[0,1]}(0.5,0.15^2)$, assuming that half of the maximum effect is most likely achieved around the midpoint of the dosing range. The prior for the hill parameter $\lambda$ is set to a gamma distribution $\Gamma(2.5,1.18)$, providing sufficient flexibility in the steepness of the dose-response curve around $ED_{50}$. For the hyperprior, we select a half-normal distribution $HN(\tau^{2})$ with $\tau=0.5$, and assess sensitivity to alternative choices $\tau \in \{0.05,0.1,0.5,1,2,3,4,5,6\}$ in the Supplemental Material.

When historical data is available (see Scenarios 1--3), we assume the between-trial prognostic heterogeneity $r \sim N(0,0.5^2)$, which provides sufficient flexibility to capture plausible variability in the prognostic effect between the trials. For between-trial predictive heterogeneity, we assume $a \sim N_{[1/3,3]}(1,0.2^2)$, reflecting expected variations in the predictive effect between the trials. To evaluate the benefit of historical borrowing, we compare the performance of SEMAP-curvature with and without incorporating historical data. When historical data is unavailable (see Scenario 4), we benchmark SEMAP-curvature against two alternative methods: LiMAP-curvature, which adopts a hyperprior $HN(3^{2})$ as recommended by \citet{han2022} for cases with limited prior knowledge, and MCP-Mod, which uses a fixed set of candidate models, including linear, emax1, emax2, exponential1, quadratic and logistic1 (corresponding to the top six figures of Figure~\ref{fig:311}).


\subsection{Results of the simulation study}
\label{sec:33}
We present the operating characteristics of SEMAP-curvature for trials simulated with the most challenging heterogeneity conditions ($r=0.2$ and $a=0.8$), representing the presence of heterogeneity in both prognostic and predictive effects between the current and historical trials. Results for other combinations of $r$ and $a$ are provided in Supplemental Material, File~\href{run:./LH2023_Supplemental_Material.pdf}{S2}.

\subsubsection{Performance in detecting the dose-response signal}
\label{sec:331}
To evaluate power in detecting dose-response signals, we generate receiver operating characteristic (ROC) curves for all true dose-response models. Each ROC curve plots the true positive rate against the false positive rate across a range of critical values, with performance improving as the curve approaches the top-left corner. This approach provides a comprehensive assessment of discriminative ability beyond reliance on a single threshold.

\begin{figure}[!ht]
	\centering
	\includegraphics[width=\linewidth]{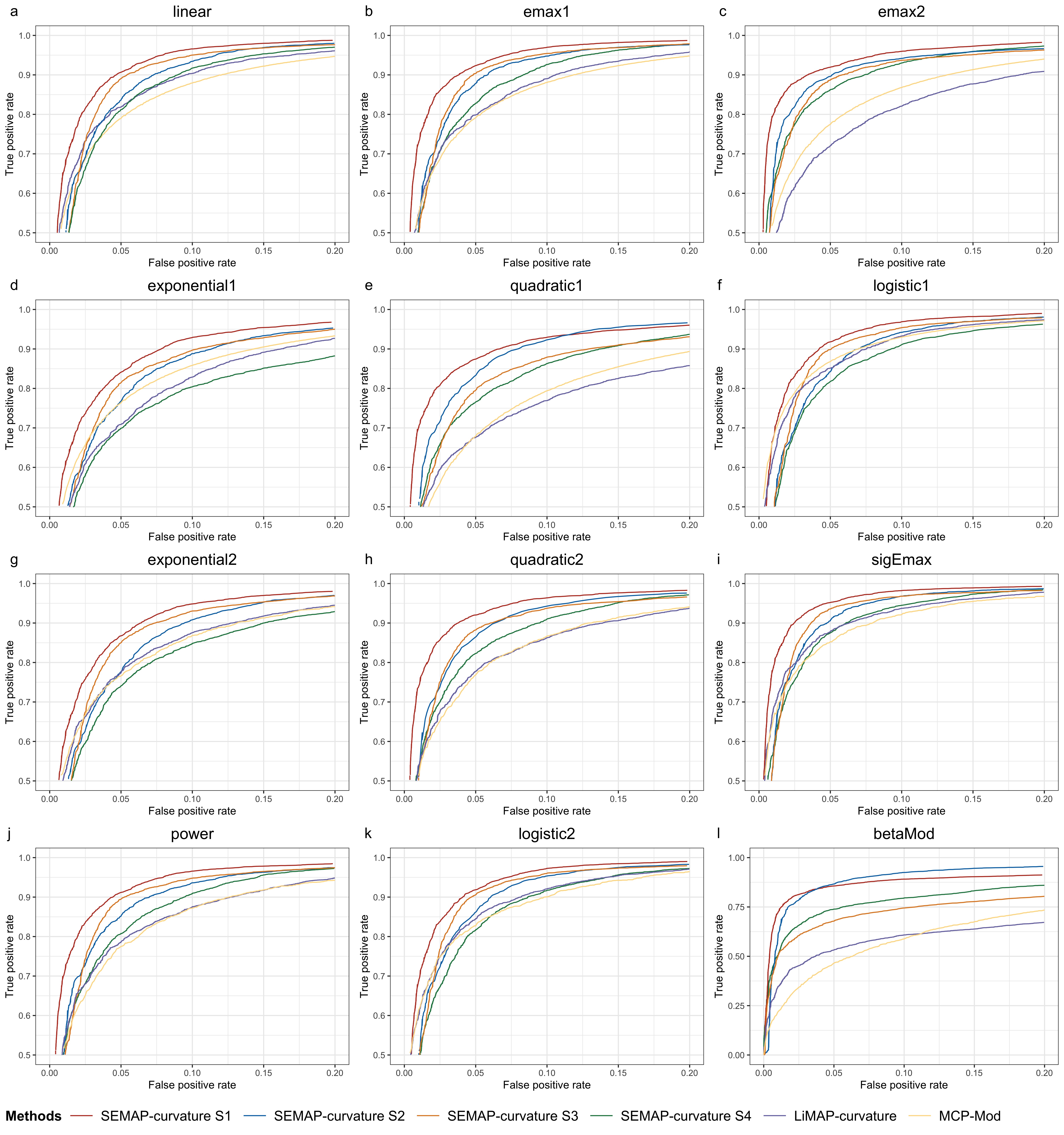}
	\caption{ROC curves of SEMAP-curvature, LiMAP-curvature and MCP-Mod across different true dose-response models with a random effect of $r=0.2$ and an effect ratio of $a=0.8$.} 
	\label{fig:321}
\end{figure}


In the absence of historical data, we compare the performance of SEMAP-curvature (S4), LiMAP-curvature, and MCP-Mod across 12 true dose-response relationships (Figure~\ref{fig:321}). SEMAP-curvature performs particularly well when the true relationship exhibits a concave downward shape, such as in the emax, quadratic, power and betaMod models, which align closely with the characteristics of the sigmoid Emax model. In these cases, SEMAP-curvature outperforms the alternatives, demonstrating strong ability to capture complex nonlinear patterns. By contrast, SEMAP-curvature underperforms LiMAP-curvature for curves such as exponential1, logistic1, and exponential2, which display an initial concave upward shape, possibly due to the prior settings for the Hill coefficient and the limited information available at early dose levels.
When the true curve is among MCP-Mod’s candidate models (Figures~\ref{fig:321}a--f), SEMAP-curvature demonstrates comparable or superior performance in models with an initial concave curvature (e.g., emax1, emax2, and quadratic1), with power gains of about 3-15\% over LiMAP-curvature and 2-14\% over MCP-Mod at a type I error rate of 5\%. When the true curve is not included in MCP-Mod’s candidate set (Figures~\ref{fig:321}g--l), SEMAP-curvature generally performs better, particularly when the true model deviates greatly from the candidate set. For example, in quadratic2 and betaMod, it achieves power gains of 8-40\% over MCP-Mod at a type I error rate of 5\%.

When historical data are available, SEMAP-curvature achieves notable gains in true positive rates, even under heterogeneity in both prognostic and predictive effects. In Scenario 1, where all dose levels overlap between historical and current trials, SEMAP-curvature (S1) delivers the best performance, with ROC curves consistently above those of other scenarios. In Scenario 2, with four overlapping doses, increased prognostic heterogeneity results in reduced performance relative to Scenario 1, though SEMAP-curvature (S2) still outperforms SEMAP-curvature (S4) without borrowing. Interestingly, Scenario 3, which includes only three overlapping doses, often surpasses Scenario 2 because all three historical doses are represented in the current trial, thereby mitigating the influence of non-overlapping doses. An exception arises in the beta model, where SEMAP-curvature (S3) performs worse than SEMAP-curvature (S4). In this setting, Scenario 3 includes only the placebo and two high doses, while the excluded intermediate doses capture the peak response of the beta model, leading to substantial information loss. These results highlight that the benefit of historical borrowing depends not only on the amount of overlap but also on whether the overlapping doses capture the most informative regions of the dose-response curve.

\subsubsection{Performance in estimating the dose-response relationship}
\label{sec:332}
\begin{figure}[!ht]
	\centering
	\includegraphics[width=\linewidth]{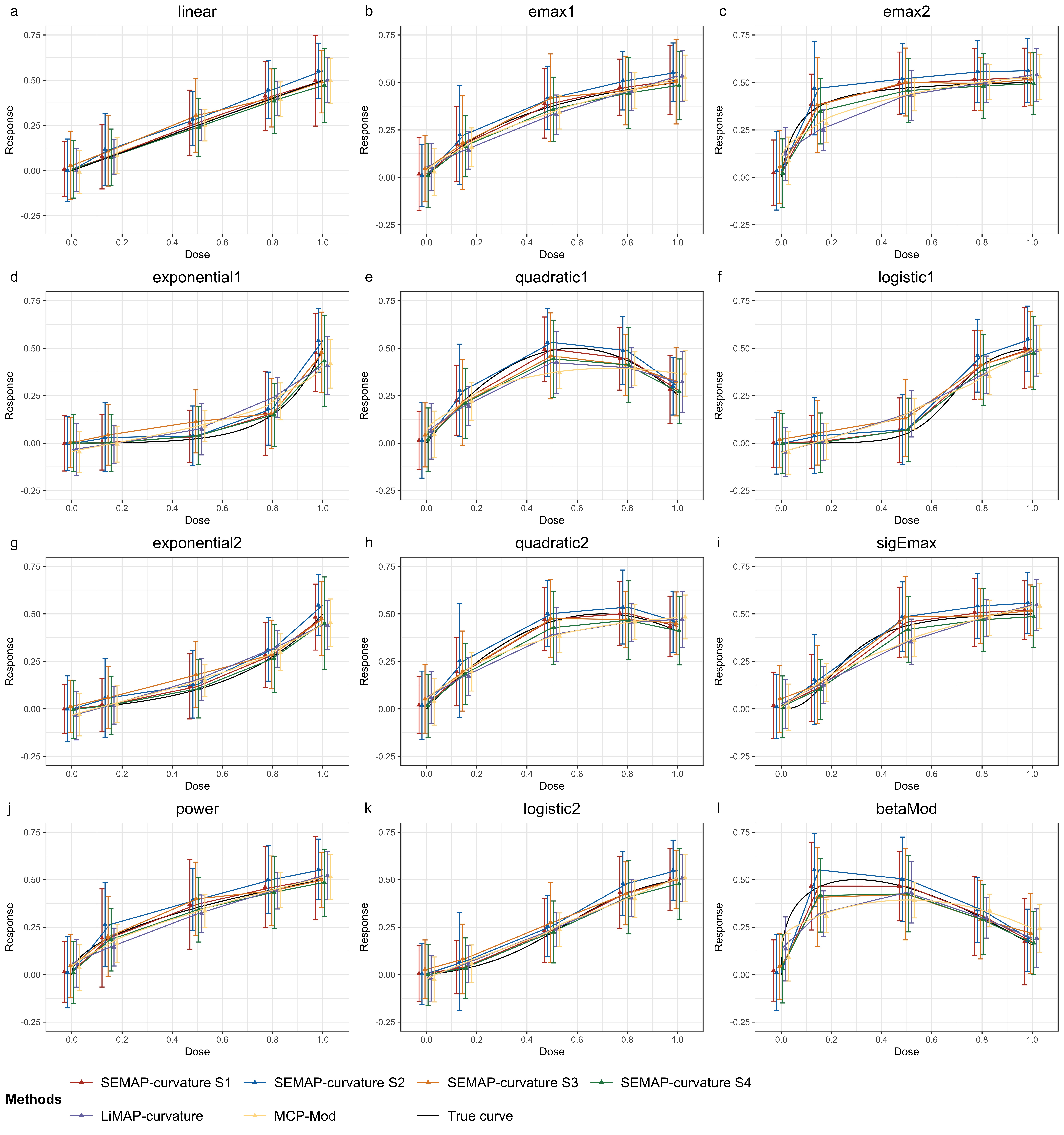}
	\caption{Dose-response curves estimated through SEMAP-curvature, LiMAP-curvature and MCP-Mod across different true underlying dose-response models with a random effect of $r=0.2$ and an effect ratio of $a=0.8$.} 
	\label{fig:322}
\end{figure}

Figure~\ref{fig:322} compares the averaged dose-response curve estimates from SEMAP-curvature (Scenarios 1--4), LiMAP-curvature and MCP-Mod across various true underlying models. Error bars represent the standard error at each dose level, showing both accuracy and precision. 

Without historical data, SEMAP-curvature (S4) aligns closely with the true curves, particularly for complex patterns such as emax2, exponential2 and logistic2. However, its relatively wide error bars indicate greater variability, suggesting that SEMAP-curvature is accurate but more sensitive to data and prior specifications. By contrast, LiMAP-curvature and MCP-Mod yield narrower error bars, suggesting more stable estimates, but their fitted mean responses deviate further from the truth in models with strong curvature, highlighting a trade-off between stability and accuracy.

When historical data are incorporated, SEMAP-curvature (S1) yields mean estimates that align more closely with the true dose-response curve than SEMAP-curvature (S4). Despite the additional information, the error bars show little reduction, probably due to the extra variability arising from prognostic and predictive heterogeneity. SEMAP-curvature (S2) also performs well, with the added non-overlapping dose helping to capture the overall curve shape, although at the expense of slight overestimation of response levels. In contrast, SEMAP-curvature (S3) performs less well, particularly for models with an initial concave upward shape such as exponential1, logistic1, and exponential2. Here, overlap is limited to placebo and two high doses, leaving the intermediate region poorly informed and reducing estimation accuracy. These results underscore that the benefit of historical borrowing depends not only on the amount of overlap but also on whether the overlapping doses cover the most informative regions of the curve.

\subsubsection{Performance in estimating the minimum effective dose}
\label{sec:333}
\begin{figure}[!ht]
	\centering
	\includegraphics[width=\linewidth]{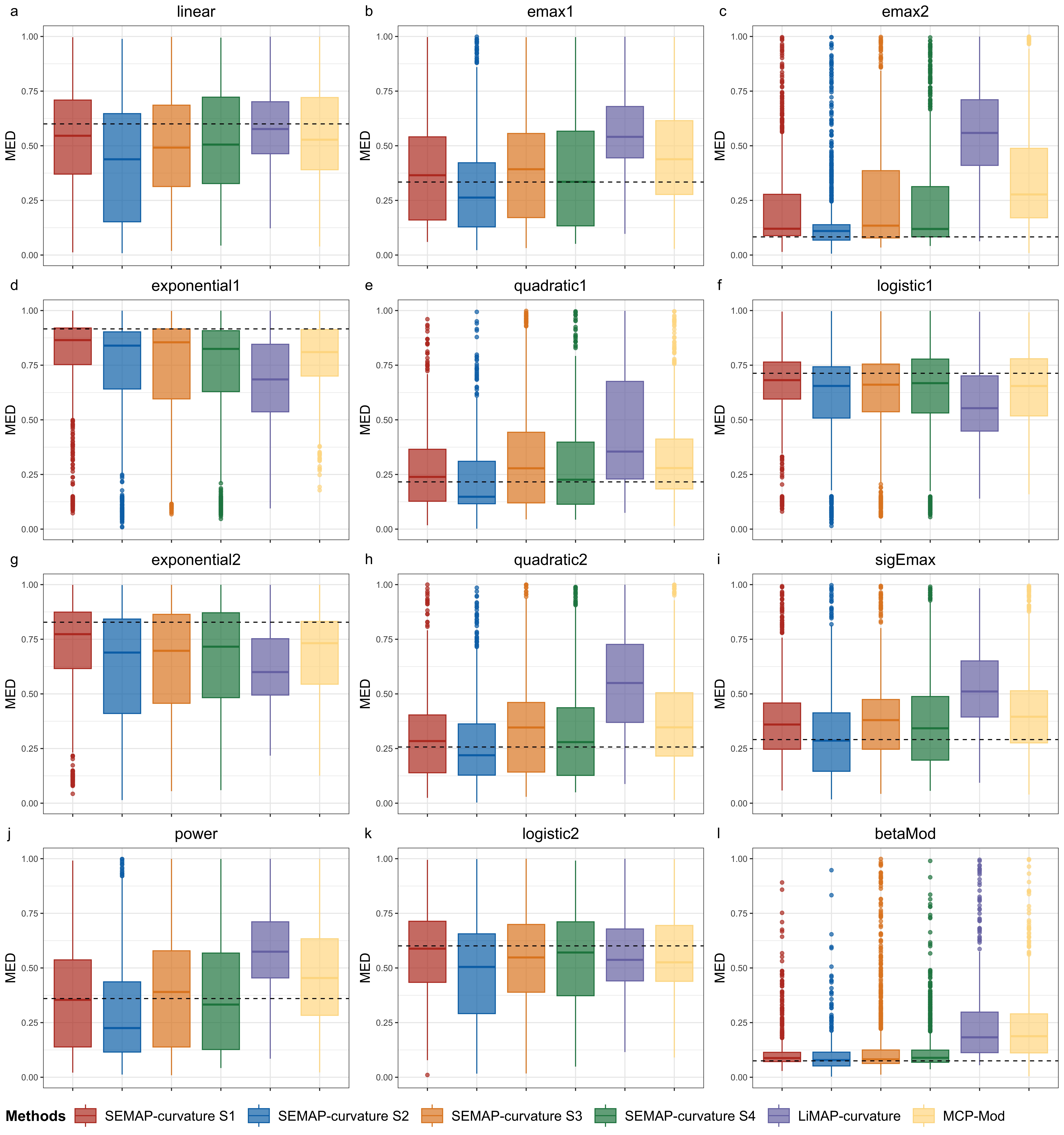}
	\caption{Boxplot distributions of estimated MED with SEMAP-curvature, LiMAP-curvature and MCP-Mod across different true underlying dose-response models with a random effect of $r=0.2$ and an effect ratio of $a=0.8$. The horizontal dashed line indicates the true MED.} 
	\label{fig:323}
\end{figure}

Figure~\ref{fig:323} presents boxplots of the estimated MED across methods and true underlying curves, with corresponding bias and mean squared error (MSE) summarised in Table~\ref{tab:323}. Both the figure and the table indicate that the accuracy and precision of MED estimates depend strongly on the underlying dose-response profile.

\afterpage{%
	\begin{landscape}
		\begin{table}[]
			\resizebox{\columnwidth}{!}{%
				\begin{tabular}{cccccccccccccc}
					\hline
					Model        & True MED & \multicolumn{8}{c}{SEMAP-curvature}                                                                         & \multicolumn{2}{c}{LiMAP-curvature} & \multicolumn{2}{c}{MCP-Mod} \\ \cline{3-14} 
					             &          & \multicolumn{2}{c}{S1}           & \multicolumn{2}{c}{S2} & \multicolumn{2}{c}{S3} & \multicolumn{2}{c}{S4} &                   &                 &               &             \\ \cline{3-10}
					             &          & \multicolumn{1}{c}{Bias} & MSE   & Bias       & MSE       & Bias       & MSE       & Bias       & MSE       & Bias              & MSE             & Bias          & MSE         \\ \hline
					linear       & 0.600    & -0.061                   & 0.060 & -0.156     & 0.093     & -0.101     & 0.072     & -0.093     & 0.077     & -0.012            & 0.029           & -0.038        & 0.047       \\
					emax1        & 0.334    & 0.051                    & 0.055 & -0.021     & 0.051     & 0.062      & 0.062     & 0.040      & 0.064     & 0.241             & 0.093           & 0.124         & 0.063       \\
					emax2        & 0.083    & 0.126                    & 0.055 & 0.071      & 0.033     & 0.170      & 0.083     & 0.140      & 0.066     & 0.465             & 0.275           & 0.260         & 0.121       \\
					exponential1 & 0.916    & -0.121                   & 0.057 & -0.215     & 0.137     & -0.188     & 0.103     & -0.196     & 0.108     & -0.233            & 0.086           & -0.129        & 0.043       \\
					quadratic1   & 0.216    & 0.055                    & 0.035 & 0.009      & 0.026     & 0.102      & 0.062     & 0.054      & 0.041     & 0.238             & 0.125           & 0.108         & 0.047       \\
					logistic1    & 0.713    & -0.043                   & 0.033 & -0.127     & 0.076     & -0.093     & 0.054     & -0.090     & 0.059     & -0.124            & 0.044           & -0.067        & 0.034       \\
					exponential2 & 0.828    & -0.121                   & 0.064 & -0.222     & 0.133     & -0.195     & 0.105     & -0.183     & 0.102     & -0.204            & 0.070           & -0.134        & 0.054       \\
					quadratic2   & 0.257    & 0.060                    & 0.041 & 0.001      & 0.031     & 0.093      & 0.058     & 0.063      & 0.052     & 0.280             & 0.129           & 0.120         & 0.056       \\
					sigEmax      & 0.291    & 0.087                    & 0.043 & 0.022      & 0.034     & 0.094      & 0.049     & 0.087      & 0.056     & 0.236             & 0.090           & 0.126         & 0.053       \\
					power        & 0.360    & 0.028                    & 0.061 & -0.055     & 0.057     & 0.036      & 0.063     & 0.016      & 0.071     & 0.227             & 0.087           & 0.120         & 0.069       \\
					logistic2    & 0.601    & -0.038                   & 0.046 & -0.122     & 0.076     & -0.073     & 0.059     & -0.071     & 0.064     & -0.032            & 0.028           & -0.037        & 0.038       \\
					betaMod      & 0.075    & 0.040                    & 0.010 & 0.018      & 0.006     & 0.072      & 0.036     & 0.052      & 0.018     & 0.170             & 0.066           & 0.142         & 0.042       \\ \hline
				\end{tabular}%
			}
			\caption{Bias and MSE of MEDs estimated through SEMAP-curvature, LiMAP-curvature and MCP-Mod across different true underlying dose-response models with a random effect of $r=0.2$, an effect ratio of $a=0.8$ and the clinical relevance threshold $0.3$.}
			\label{tab:323}
		\end{table}
	\end{landscape}
}

In the absence of historical data, certain dose-response models, such as exponential1, exponential2, logistic1 and sigEmax produce more biased MED estimates across all three approaches. Among these, SEMAP-curvature (S4) generally delivers the least biased estimates for nonlinear shapes, with medians often aligning with the true MED (horizontal dashed line), although its precision varies across models. For profiles such as emax2, quadratic1, quadratic2 and betaMod, SEMAP-curvature performs particularly well, achieving both low bias and high precision, while for profiles such as emax1 and power, it maintains accurate median estimates but with greater variability, reflecting reduced precision. LiMAP-curvature performs well in simpler settings, like the linear model, but exhibits larger bias and higher MSE in highly curved patterns, including emax2 and quadratic2, indicating difficulty in capturing complex dose-response profiles. MCP-Mod also tends to show larger bias in nonlinear relationships, such as emax1 and quadratic1, underscoring its limitations when the true curve departs from its candidate model set.

When historical data are incorporated, SEMAP-curvature shows clear gains in the accuracy and precision of MED estimates. These improvements are most evident when dose levels in the current and historical trials fully or largely overlap, leading to reduced bias and variability. For example, in Scenario 1 with complete overlap, SEMAP-curvature produces minimal bias and the narrowest interquartile ranges across most models. In Scenario 2, the inclusion of one non-overlapping dose introduces some variability, yet SEMAP-curvature still outperforms Scenario 4, yielding lower bias and MSE. Scenario 3 also improves upon Scenario 4, particularly when the overlapping doses capture the most informative region of the dose–response curve. However, in profiles such as quadratic1 and quadratic2, where high-response doses are absent from the overlap, the benefit of historical borrowing diminishes and accuracy may even decline relative to Scenario 4.


\section{Discussion}
\label{sec:4}
In the present work, we have introduced MAP-curvature, a general and flexible model-free framework for dose-response modelling that permits the choice of different default dose-response functions depending on the characteristics of the relationship under investigation. The proposed method balances adaptability and accuracy by imposing smoothness on the curve while allowing complexity to vary with the selected default model. LiMAP-curvature \citep{han2022}, which assumes a linear default dose-response function, is well suited to scenarios where a straight line approximation properly captures the dose-response. Its simplicity and computational efficiency make it practical for early-stage dose-finding studies with limited prior knowledge. In contrast, SEMAP-curvature, the focus of this study, adopts the sigmoid Emax model as the default dose-response function. Such a nonlinear relationship is usually observed in pharmacological research and provides greater flexibility for modelling complex dose-response patterns. A key advantage of the MAP-curvature framework is its adaptability: users are not restricted to a single default dose-response model but can instead choose the most appropriate one based on prior knowledge or emerging data.

Our extensive simulations have demonstrated the superior performance of SEMAP-curvature in detecting dose-response signals, estimating dose-response curves and identifying target doses. Across a range of dose-response relationships, particularly those with concave downward shapes resembling the sigmoid Emax model, SEMAP-curvature generally outperforms alternative methods, including LiMAP-curvature and MCP-Mod, in establishing PoC and in estimating both the dose-response curve and the MED. Its advantage in MED identification is notably pronounced, yielding estimates with lower bias and reduced MSE. Although SEMAP-curvature shows greater variability, reflected in wider error bars, this does not materially compromise its overall accuracy or reliability across different models when estimating responses at various doses.

Furthermore, extending MAP-curvature within a Bayesian hierarchical framework allows the incorporation of historical data, which is particularly valuable in early-stage trials with limited sample sizes. The framework automatically calibrates the degree of information borrowing based on the heterogeneity between historical and current trials, which ensures robustness even when prognostic and predictive effects differ across studies. Our simulation study has demonstrated that historical borrowing increases the power to detect dose-response signals and improves the precision of dose-response estimates, particularly when there is substantial overlap in dose levels between trials.

The practical application of SEMAP-curvature in dose-finding trials requires careful attention to the choice of the hyperparameter $\tau$. Our simulation study has shown that smaller values of $\tau$ (\textit{e.g.}, $\tau=0.1$) perform well when the true dose-response curve closely resembles a sigmoid Emax shape, as they impose a stronger prior belief in sigmoidal patterns and increase sensitivity to such relationships. In contrast, when the dose-response curve greatly deviates from a sigmoid Emax shape (\textit{e.g.}, the exponential1 model), larger values of $\tau$ offer greater flexibility to capture additional complexity and variability in the data. Therefore, the selection of $\tau$ should be guided by prior knowledge about the underlying dose-response relationship. To balance robustness and flexibility, our sensitivity analysis (see Supplemental Material, File~\href{run:./LH2023_Supplemental_Material.pdf}{3}) suggests moderate values (\textit{e.g.}, $\tau=0.5$ or $1$) as practical defaults, ensuring reliable and accurate estimation across a wide range of scenarios.

While SEMAP-curvature offers notable advantages, it also has limitations. Its performance depends critically on the choice of the hyperparameter $\tau$ and the specification of prior distributions, both of which require careful consideration and, ideally, prior knowledge of the expected dose-response relationship. Future work will aim to refine SEMAP-curvature to overcome these challenges and to extend its application to more complex adaptive trial designs, where doses are updated dynamically as data accrue. In addition, integrating SEMAP-curvature with advanced dose-finding algorithms and real-world data could further enhance its utility, particularly in the context of personalised medicine.

Although this work focuses on the presentation and evaluation of SEMAP-curvature incorporating a single historical trial, its extension to borrowing information from multiple historical trials or parallel trials is readily implementable. Expanding SEMAP-curvature to accommodate longitudinal data and other common endpoint types would significantly enhance its applicability across a variety of scenarios. Additionally, the development of user-friendly software packages or modules implementing SEMAP-curvature would facilitate its broader adoption and accessibility within the clinical research community, promoting its practical utility in dose-finding studies.



\subsection*{Disclosure}
The opinions expressed in this article are solely those of the authors and not those of their affiliations, and the authors' affiliations bear no responsibility for the accuracy or reliability of the information presented.

\subsection*{Conflict of interest}
The authors declare no conflicts of interest.






\bibliography{LH2023_Manuscript}

\end{document}